# Deep Learning-enabled Virtual Histological Staining of Biological Samples

Bijie Bai [†,1,2,3], Xilin Yang[†,1,2,3], Yuzhu Li[1,2,3], Yijie Zhang[1,2,3], Nir Pillar[1,2,3], and Aydogan Ozcan[*,1,2,3]

[1]Electrical and Computer Engineering Department, University of California, Los Angeles, CA, 90095, USA.

[2]Bioengineering Department, University of California, Los Angeles, 90095, USA.

[3]California NanoSystems Institute (CNSI), University of California, Los Angeles, CA, USA.

[†]Equal contributing authors

[*]Correspondence: Aydogan Ozcan. Email: ozcan@ucla.edu

## Abstract

Histological staining is the gold standard for tissue examination in clinical pathology and life-science research, which visualizes the tissue and cellular structures using chromatic dyes or fluorescence labels to aid the microscopic assessment of tissue. However, the current histological staining workflow requires tedious sample preparation steps, specialized laboratory infrastructure, and trained histotechnologists, making it expensive, time-consuming, and not accessible in resource-limited settings. Deep learning techniques created new opportunities to revolutionize staining methods by digitally generating histological stains using trained neural networks, providing rapid, cost-effective, and accurate alternatives to standard chemical staining methods. These techniques, broadly referred to as *virtual staining*, were extensively explored by multiple research groups and demonstrated to be successful in generating various types of histological stains from label-free microscopic images of unstained samples; similar approaches were also used for transforming images of an already stained tissue sample into another type of stain, performing virtual stain-to-stain transformations. In this Review, we provide a comprehensive overview of the recent research advances in deep learning-enabled virtual histological staining techniques. The basic concepts and the typical workflow of virtual staining are introduced, followed by a discussion of representative works and their technical innovations. We also share our perspectives on the future of this emerging field, aiming to inspire readers from diverse scientific fields to further expand the scope of deep learning-enabled virtual histological staining techniques and their applications.

## Introduction

Over the past century, histological staining has been established as a principal tool for tissue examination in disease diagnostics and life-science research[1,2]. By labeling different biological elements with specific markers based on their biochemical properties, histological staining enables the visualization of tissue and cellular structures and allows the assessment of pathophysiology and disease development when the stained samples are observed under a light microscope[3–5]. Various stain types have been developed and routinely performed in histology labs, corresponding to the different biological features to be highlighted. For example, hematoxylin and eosin (H&E) stain creates a contrast between the nuclei and the extracellular tissue matrix and is the most frequently used stain in histopathology[6]; Masson's trichrome (MT)[7] and Periodic acid–Schiff (PAS) stains[8], two examples of special stains, highlight collagen fibers and glycoproteins, respectively, and are commonly used in cardiac and kidney pathology[4]. Immunohistochemical (IHC) staining, a more advanced molecular staining technique, highlights the presence of specific epitopes based on antigen-antibody binding and is extensively utilized in pathology[9,10].

These standard histological staining procedures are routinely carried out in pathology laboratories following a decades-old workflow in which time-consuming sample preparation (e.g., tissue fixation, embedding, and sectioning) and laborious histological staining steps are performed before the samples can be inspected under a light microscope (Fig. 1a). Such chemical-based staining procedures require designated laboratory infrastructure and manual supervision from trained histotechnologists, making them expensive and not accessible in resource-limited settings. The multi-step staining protocols accompanied by laborious manual supervision by experts result in long turnaround times (e.g., days to weeks) and consequently delay the disease diagnosis and treatment. Moreover, the destructive nature of the chemical staining process prohibits additional staining and further molecular analysis on the same section. As another disadvantage, the toxic chemical compounds involved in the staining process generate significant amounts of waste and consume >1 million liters of water globally. All in all, there is a strong demand for alternative staining methods that can provide rapid, cost-effective, and accurate solutions to overcome these limitations.

In recent years, the wide adoption of digital pathology[11,12], combined with the ever-growing deep learning techniques, has created new opportunities to revolutionize these decade-old staining methods. Deep learning-based image transformations from faster, simpler, and easier-to-access microscopy modalities to more advanced but difficult-to-obtain ones, have been extensively studied for various biological samples[13–17]. As an extension of this line of research in the domain of histopathology, deep learning-based methods have been developed to virtually replicate the images of chemically stained slides using only the microscopic images of unlabeled samples[18,19], eliminating the need for chemical staining procedures (Fig. 1b). These virtual staining methods were demonstrated to successfully generate different types of histological stains using various label-free imaging modalities, such as autofluorescence imaging and quantitative phase imaging (QPI)[18,20]. The principal idea of using deep learning-based image transformation to bypass the traditional histological staining process also enables the transformation from one existing stain type into another[19,21] (Fig. 1b). Stain-to-stain transformation methods can provide pathologists with additional types of stains in an instant manner, without changing their current workflow. These new technologies not only reduce costs, labor and time-to-diagnosis, but also open up new possibilities for stain multiplexing and *in-vivo* staining[22,23], greatly expanding the field of histopathology beyond what is possible within the traditional chemical staining paradigm currently in use.

In this Review, we provide a comprehensive overview of recent advances in deep learning-enabled virtual histological staining techniques. We will first introduce the basic concepts and typical workflow of virtual staining via deep learning. Next, we will highlight some key results from representative works in this field and dive into their technical details, including the innovative aspects of their data preparation and network training methods. We group these approaches into (i) label-free virtual staining (Table 1) and (ii) stain-to-stain transformations (Table 2), where the former computationally generates the virtual histological images from label-free images captured using unstained samples, and the latter digitally transforms the images of already stained tissue samples (e.g., H&E-stained) into other types of stains (e.g., MT and IHC). Finally, we will share our perspectives on the future directions in this rapidly evolving virtual staining field, also shedding light on areas that need further research effort. We believe this Review will serve as an atlas of the technical developments in this research area, which can introduce the top-level concepts and the up-to-date research progress to scholars who are relatively new to this field. We also hope this Review will be of broad interest to optical engineers, microscopists, computer scientists, biologists, histologists and pathologists, providing an introduction to virtual staining technologies and the transformative opportunities these approaches can create in histopathology.

**Development of a virtual staining model**

The workflow needed to develop a label-free virtual staining or a stain-to-stain transformation model typically consists of image data collection, image pre-processing, as well as network training and validation, as shown in Figs. 2-3. Depending on the learning schemes used for creating the virtual staining models (e.g., supervised or unsupervised), the corresponding upstream data collection and pre-processing methods will differ. In supervised training settings, perfectly cross-registered input and ground truth image pairs are needed for training an image transformation virtual staining network. Therefore, multi-stage image registration (Fig. 2a) or pre-trained data generation models (Fig. 3a) are usually required to generate well-matched training images. On the other hand, in the unsupervised training settings, the images from the input and ground truth domains are not necessarily paired (see Fig. 2b and Fig. 3b). This saves effort in data pre-processing, however, increases the complexity of the network architecture and the training schedule. Cycle-consistency-based learning frameworks (e.g., CycleGANs[24]), are commonly used in unsupervised training scenarios, which learn to map the distribution of the input images to the ground truth domain, matching the color and contrast.

For both learning schemes, developing a reliable virtual staining model often involves acquiring and processing a large volume of data and carefully designing and training the neural networks (see Figs. 2a-b and Figs. 3a-b), which could take a substantial amount of time. However, this model development stage is a one-time process; this is, in principle, very similar to the development and fine-tuning of the protocols of a histochemical staining workflow that involves various chemical optimization steps, all of which also constitute a one-time development effort. Once a satisfactory virtual staining model is obtained and validated, its blind inference is rapid and repeatable (Fig. 2c and Fig. 3c), which only takes a few minutes to create a whole-slide virtual histological image of a tissue section using a standard computer, without waiting for any chemical staining procedures to be completed. This virtual staining process not only saves time and labor, but also eliminates the use of toxic staining compounds, and is, therefore, environmentally more friendly.

**Label-free virtual staining**

The use of deep learning to successfully achieve virtual staining of label-free tissue samples using autofluorescence images was demonstrated by Rivenson *et al.*[18,25], in which deep neural networks were trained to transform the images of various unstained tissue sections, e.g., salivary gland, thyroid, liver, and lung, into multiple histological stains including H&E, MT, and Jones silver stain, closely matching the bright-field images of the same tissue sections after the standard histochemical staining (Fig. 4a). Over the last few years, several studies have been carried out to further expand this label-free virtual staining technique[26–28]. As summarized in Table 1 and Figure 4, multiple types of histological stains were successfully replicated using different image contrast mechanisms on various types of samples, which greatly enriched the application areas of virtual staining methods. Furthermore, by adding customized digital staining matrices to the autofluorescence images and using their combination as the neural network input, Zhang *et al*. achieved micro-structured and multiplexed histological stains on the same tissue section with a single network, which is not feasible with the traditional histochemical staining workflow[22] (Fig. 4b). In fact, the autofluorescence emission signatures of biological tissue carry convoluted spatial-spectral information of its metabolic state and pathological condition[29,30]. Therefore, in addition to the standard histochemical stains such as H&E and MT, the autofluorescence images of label-free tissue can be utilized to generate more complex molecular stains, e.g., highlighting a specific protein expression, as currently done by conventional IHC staining protocols commonly employed in histology labs. For example, Bai *et al*. successfully demonstrated virtual IHC staining of human epidermal growth factor receptor 2 (HER2) protein using the autofluorescence images of unlabeled breast tissue sections[31] (Fig. 4c), significantly extending the reach of virtual tissue staining via label-free autofluorescence imaging.

Though powerful, autofluorescence microscopy is not the only imaging modality that enables label-free virtual staining. Several different imaging modalities that bring contrast for unlabeled biological samples have been explored for virtual staining. For example, QPI, which is based on the refractive index distribution of unstained biological samples, was also utilized for virtual staining. Rivenson *et al*. used the quantitative phase images of various label-free tissue sections and transformed them into virtual H&E, Jones, and MT stains using convolutional neural networks, matching their histochemically-stained counterparts in terms of staining quality[20] (Fig. 4d). In another work, Nygate *et al*. demonstrated the virtual staining of human sperm cells using QPI, allowing fertility evaluation in real-time[32]. QPI using oblique back-illumination microscopy was also utilized by Abraham *et al*. to generate virtual H&E staining of thick and intact mouse brain samples[33].

Other microscopy methods, such as nonlinear optical imaging, have also been adopted for label-free virtual staining. Borhani *et al*. used two-photon excitation fluorescence (TPEF) alongside fluorescence lifetime imaging (FLIM) as the network input to virtually stain rat liver samples with H&E[34]. Pradhan *et al*. combined coherent anti-Stokes Raman scattering (CARS), second-harmonic generation (SHG) microscopy and TPEF to create virtual H&E staining on human colon samples (Fig. 4e)[35]. Some additional label-free imaging methods were also applied for deep learning-based virtual staining tasks. To list some examples, bright-field imaging of unstained carotid artery sections was used to generate multiple types of stains, such as H&E and picrosirius red (PSR) (Fig. 4f)[36,37]; multichannel total absorption photoacoustic remote sensing (TA-PARS) was utilized to create virtual H&E staining of human skin tissue (Fig. 4g)[38]; UV imaging of whole blood smears was computationally transformed into Giemsa staining[39]; UV photoacoustic microscopy was also demonstrated to achieve virtual H&E staining of mouse brain [40] and frozen sections of bone tissue[41]. As another example, Mayerich. *et al.*, developed a shallow artificial neural network (ANN) model, without any hidden layers, to learn a pixel-to-pixel mapping from Fourier transform infrared (FT-IR) spectroscopy to bright-field imaging, targeting multiple stains on human breast tissue[42]; in this approach, however, the 2D spatial information of the label-free image is ignored, and the virtual staining is performed using the spectrum at each pixel individually, i.e., separate from other pixels. Due to the lack of deeper convolutional layers that process the 2D texture information of tissue structure, such a one-dimensional approach presents limited staining performance and generalization[42].

Virtual staining techniques can also be integrated with noninvasive microscopic imaging modalities, achieving *in vivo* virtual staining without a biopsy (i.e., "virtual biopsy"). As demonstrated by Li *et al*.[23], an *in vivo* virtual staining method using reflectance confocal microscopy (RCM) can be used to create virtual H&E staining of human skin tissue (Fig. 4h), which can potentially be used for rapid diagnosis of malignant skin neoplasms while eliminating unnecessary biopsies and scars as well as cumbersome sample preparation steps.

**Stain-to-stain transformations**

Deep learning also enables the transformation of the microscopic images of an already stained tissue into other types of stains, providing additional contrast information for differentiating, e.g., various cellular structures and helping improve the diagnosis. For example, Gadermayr *et al*. demonstrated stain-to-stain transformations using deep learning, achieving image transformations from PAS stain into Acid Fuchsin Orange G (AFOG), CD31 IHC, and Collagen III (Col3) stains[43]. These stain-to-stain transformations allowed them to compare the segmentation accuracy of glomeruli under different stain types within the same field-of-view (FOV), which is not possible with standard histology since a given tissue section can, in general, be stained with only one type of stain. Additional applications of deep learning-based stain-to-stain transformations were demonstrated over the last several years, as summarized in Table 2 and Figure 5.

Stain-to-stain transformations offer a highly convenient and rapid approach to generating stain types that are more difficult to obtain using more prevalent and cheaper stains (such as H&E). For instance, a default choice of the "source" stain used as the input for stain-to-stain transformations is the H&E stain due to its wide accessibility and cost-effectiveness[6]. The transformation of the H&E stain into special stains, which are used to visualize particular tissue structures not revealed by H&E staining[4], was demonstrated by several research groups[21,44–53]. In the work of de Haan *et al*. (Fig. 5a)[21], deep neural networks were trained to transform H&E-stained human kidney samples into special stains, including Jones silver, MT, and PAS stains. The stain-to-stain transformation improved the diagnostic accuracy in a blinded study and will be transformative in reducing the turnaround time of the inspection of non-neoplastic kidney biopsies. As another example, Levy *et al*. generated virtual trichrome stain from H&E staining of human liver samples to study the staging of liver fibrosis[44]. Furthermore, Lin *et al*. demonstrated multiplexed stain transfer from H&E to PAS, MT, and Periodic Schiff-Methenamine (PASM) stains on human kidney samples[45].

Other than special stains, different IHC-based stains were also successfully generated using H&E images as input. Liu *et al*. demonstrated the transformation from H&E into IHC staining of Ki-67 on neuroendocrine and breast tissue samples (Fig. 5b)[46]. Their virtually generated IHC images showed a high degree of agreement with the ground truth IHC images on both Ki-67 positive and Ki-67 negative cells. Xie *et al*. achieved virtual IHC staining of cytokeratin 8 (CK8) from H&E-stained 3D whole prostate biopsy samples, potentially improving the risk stratification of prostate cancers[47]. Other virtual IHC stains generated/transformed from H&E include HER2[48,49] and Phosphohistone H3 (PHH3)[50] on breast samples, SOX10 on liver samples[51], and Cytokeratin (CK) on liver[52] and stomach[53] samples.

Compared to colorimetric IHC staining that uses chromatic markers to highlight specific antibodies attached to their target ligands, immunofluorescence (IF) staining, also based on antigen recognition elements, allows for improved sensitivity and signal amplification[54] by using fluorescent labels. The generation of virtual IF staining from other stain types was also reported: Ghahremani *et al*. used Ki-67 IHC stained images to generate multiplexed virtual IF staining of various biomarkers on human lung and bladder samples (Fig. 5c)[55]. Burlingame *et al*. achieved the transformation of H&E-stained images into virtual IF-stained images of pan-cytokeratin (panCK) biomarker on human pancreatic cancer samples[56] (Fig. 5d).

Besides performing stain-to-stain transformations using the formalin-fixed, paraffin-embedded (FFPE) tissue sections, the generation of H&E staining from ultraviolet surface excitation microscopy (MUSE) images of the Hoechst stained fresh mouse brain was also reported (Fig. 5e)[57] as another form of stain-to-stain transformation, with the additional advantage that Hoechst staining is very fast and relatively simple.

**Training data preparation**

The training of the aforementioned virtual staining models usually requires image data collected from both the input and the target (ground truth) domains so that the model can be trained to exploit and translate the information from the input domain to the target domain. Between the collection of raw image data and the training of the virtual staining models, image pre-processing steps are necessary to prepare the datasets for successfully learning the image transformation. These data pre-processing steps mainly focus on cross-registering the input and target image pairs, which is essential for supervised learning frameworks and eliminating unexpected outliers, such as misaligned image pairs and staining artifacts[18]. Another aim of data pre-processing is to address the domain shift problem[58], which refers to the statistical distribution deviation within a model's training dataset or between the training dataset and a dataset it encounters during testing. Such a deviation might originate from several sources, e.g., the variations in the image acquisition set-up and/or the staining variations due to the nature of the chemical-based tissue staining workflow. Using proper data normalization methods, such deviations among the images could be minimized, so that the statistical distribution of the image data is confined within a certain range/domain to promote the learnability of the virtual staining tasks[59,60].

The cross-registration of the input and target image pairs is commonly adopted in supervised learning-based frameworks. An example of such a registration process was reported in the work of Rivenson *et al.* on virtual staining of autofluorescence images, where a multi-model image registration algorithm was implemented. This algorithm starts with a coarse registration of the autofluorescence images of the label-free tissue section with respect to the bright-field images of the same tissue sections after the corresponding histochemical staining process was completed, where the roughly matched FOVs of both imaging modalities were extracted by searching for the highest cross-correlation score[18]. Then an affine transformation was estimated by matching the feature vectors (descriptors) between the extracted histologically stained images and the autofluorescence images, which was then applied to the stained images to correct any changes in scale or rotation. In the last, finer image registration step, a virtual staining network was first trained through a low number of iterations with the roughly matched images to learn the color mapping. Then the trained pseudo model was applied to the autofluorescence images to assist the local feature registration using an elastic pyramidal registration algorithm[61,13], which helped to achieve pixel-level co-registration accuracy between the autofluorescence images of label-free tissue sections (input images) and their corresponding histochemically stained ground truth images. Similar multiple-stage image registration algorithms were also utilized in several other supervised learning-based virtual staining methods[20–23,31].

As an alternative approach, Borhani *et al.*[34] aligned the input and target images by using a combination of scale-invariant feature transform (SIFT)[62] keypoints and random sample consensus (RANSAC) function fitting[63] at two stages that were at different length scales. After both the input and the target images were converted to comparable grayscale images, SIFT algorithm was first applied to locate the characteristic structural keypoints, each of which was described by a feature vector that characterized its neighboring profile. The detected keypoints between each image pair were then matched by employing the nearest neighbor search between their feature descriptor vectors, which is used to form a warping function between the image pairs. Next, the RANSAC algorithm was used to remove any outlying and erroneous matches, determining a specific affine transform, which was applied to the histologically stained images to co-register them to the label-free input images. Similar to this feature detection and matching method, Burlingame et al.[56] registered the H&E and IF whole-slide images (WSIs) using an affine transformation estimated from matched speeded-up robust features (SURF)[64], which were extracted from hematoxylin and DAPI binary masks of cell nuclei generated by a thresholding method[65].

For the label-free microscopic images, the domain shift problem is usually observed as imaging variations that occur under different experimental conditions. This might be caused by, e.g., different imaging hardware/settings, inconsistent image acquisition environments, and variations of the specimen characteristics or sample preparation protocols. To address this domain shift problem, image normalization is often applied to the input label-free images before feeding them into a virtual staining neural network. For instance, to avoid intensity variations caused by potential photobleaching in autofluorescence imaging, Rivenson et al. normalized the input autofluorescence images by subtracting the mean value across the entire tissue slide and dividing it by the standard deviation of the pixel values[18]. Alternatively, to mitigate these variations and enhance the image contrast, some works saturated the top 1% and the bottom 1% of the pixels[38,39]. Pradhan et al. also reported that normalizing label-free nonlinear multi-modal (NLM) images in a pixel value range from -1 to 1 could avoid large number multiplications during the training process, helping with better network convergence[35]. Besides these variations that can be mitigated by proper normalization, sometimes the captured microscopic images might be corrupted with, e.g., defocusing, motion blur, and readout errors. For example, Zhang et al. presented a virtual staining framework using defocused autofluorescence images as input, in which an autofocusing network was first trained to bring the randomly defocused (non-ideal) images into focus, followed by a virtual staining network (that is jointly trained) to generate in-focus virtually stained tissue images[66]; this was used to significantly speed up the whole slide imaging since fine autofocusing during the tissue scanning process is not needed in this case. Similarly, other non-ideal imaging conditions at the input end can also be mitigated using pre-trained neural networks[67,68].

Domain shift problems also exist in the histologically stained images, typically observed as immensely inconsistent color and contrast due to chemical staining variations (from lab-to-lab or histotechnologist-to-histotechnologist). One common method to eliminate such variations in the training dataset is to use stain-separation and color-normalization algorithms. Traditionally, these methods are implemented through color deconvolution and optical density mapping[69,70,71]. For instance, Burlingame et al. normalized the H&E images using the Macenko method to mitigate the inter-sample staining variations[56]. Recently, deep learning-based stain normalization has also been used because of its ability to take the spatial features of the tissue structure into account, avoiding improper staining that can be generated in the traditional algorithmic stain normalization methods[72,73]. Besides using normalization methods to unify the color and contrast of the chemically stained images, another direction to mitigate such domain shift problems in the ground truth is to incorporate these variations into the training dataset. For example, de Haan *et al*. used a pre-trained style transfer network to transform the H&E stained images into different styles for the training of a stain-to-stain transformation network[21], ensuring that the method is effective when applied to various styles of H&E-stained tissue samples regardless of the inter-technician, inter-lab or inter-equipment variations. In addition to these image registration and normalization processes, an algorithmic or manual data cleanup is also commonly performed to remove the undesirable data that may mislead the network training, such as deformed tissue sections or images with random dirt/dust placed on the sample.

**Network architecture and training strategies**

Various network structures have been reported for virtual staining, among which the generative adversarial network[74] (GAN) is one of the most commonly and widely used frameworks due to its strong representation capability[18,20–23,31,32,36–38,45,48,53,56,75,76]. In a GAN framework, two deep neural networks, the Generator and the Discriminator, are optimized in a simultaneous and competitive manner[77]. The Generator network learns to perform the image transformation from the input domain to the target domain, which typically utilizes the U-Net architecture[78] or its variants. On the other hand, the Discriminator network is a classifier that learns to distinguish between the virtually-stained images generated by the Generator and the target histologically stained images. During the training, the Discriminator looks at the virtually stained images and returns an adversarial loss to the Generator, helping it to generate images that cannot be distinguished by the Discriminator. When the training enters an equilibrium state, the Generator is able to create virtually stained images that cannot be differentiated from the histologically stained images by the Discriminator. However, in the standard GAN framework where the Generator is solely optimized by an adversarial loss, the resulting Generator only mimics the colors and patterns of the target images without learning the underlying correspondence between the input and the target images, resulting in severe hallucinations at the micro-scale[19]. To overcome this hallucination problem, various other pixel-wise loss functions, such as mean absolute error (MAE)[18,21,22,32,36,37,48,53,56], mean square error (MSE)[18,76], structural similarity index measure (SSIM)[31,79], Huber loss[31], reversed Huber loss[23], and color distance metrics[53] are incorporated into the Generator loss terms (in addition to the Discriminator loss) to regularize the GAN training; these additional loss terms are calculated using the virtually generated images and their corresponding ground truth (histochemically stained images). Moreover, image sparsity regularization terms such as total variation[80] were also exploited in some works to eliminate or suppress different types of image artifacts created by the Generator[18,20–22,31].

When precisely registered input and target image pairs are available, it is often the best strategy to train a virtual staining network using a supervised learning scheme since the pixel-wise loss functions listed earlier can be accurately evaluated for the optimization of the Generator. With a well-registered training dataset, a typical virtual staining network architecture that uses supervised learning is summarized in Fig. 6a. This GAN-based virtual staining framework demonstrated success for various tissue-stain combinations; however, it falls short in applications where paired images of the same tissue FOVs are hard (or even impossible) to acquire, such as the stain-to-stain transformation tasks where a given tissue section is typically stained only once, making it practically impossible to create pairs of histochemically stained images of the same sections with different types of stain. One approach used to mitigate this limitation and generate paired images with different stain types involved multiple pre-trained virtual staining networks[21], which, however, may also induce an unavoidable distribution shift between the target histological images and the output from the pre-trained networks. To overcome this dilemma, Yang *et al*. demonstrated a cascaded neural network (C-DNN)[76] architecture, where a virtual staining network $A$ was followed by a jointly optimized stain transfer network $B$ as shown in Fig. 6b. By using two groups of paired label-free, histochemically stained, and virtually stained images, the C-DNN used structural loss terms like MAE directly on histochemically stained images from both the input and output domains to improve the quality of virtual stain-to-stain transformations[76].

Unlike supervised learning, the training of virtual staining networks using unsupervised learning schemes does not require cross-registered image pairs. One of the most frequently used unsupervised learning frameworks for virtual staining is the CycleGAN[24] architecture (Fig. 6c) and its variants, which consist of two cascaded Generators trained jointly to perform the image transformations between the domain $X(x)$ and the domain $Y(y)$ in a cyclic manner. In one cyclic loop, the Generator $G$ first performs the transformation from domain $X(x)$ to domain $Y(G(x))$, followed by the Generator $F$ performing the transformation from domain $Y(G(x))$ back to domain $X$ $(x^* = F(G(x)))$. Similarly, a symmetric transformation from domain $Y$ $(y)$ to domain $X$ $(F(y))$ and then back to domain $Y$ $(y^* = G(F(y)))$ is accomplished in another cyclic loop. Cycle-consistency losses such as MAE[26,44,46,57], MSE[35,52], and SSIM[46,57] are typically used in such a CycleGAN training framework to measure the differences between $x \leftrightarrow x^*$ and $y \leftrightarrow y^*$. Moreover, an adversarial loss is also applied on $x \leftrightarrow F(y)$ and $y \leftrightarrow G(x)$ to ensure the generation of realistic images. In addition to the cycle-consistency losses and the adversarial loss terms, the perceptual embedding consistency (PCE) loss between the latent features extracted by the encoders of the two Generators was used by Lahiani *et al*.[52] and Liu *et al*.[46] to improve the virtual staining performance further. When the training of CycleGAN converges, the Generator $G$ is able to transfer the images from domain $X$ to domain $Y$ while the Generator $F$ can inversely perform the transformation from domain $Y$ to domain $X$. Either of these Generators can be taken out and used in the inference phase depending on the desired virtual staining task. One notable issue of using CycleGANs to perform virtual staining tasks is the intensity mismatch; for example, label-free input images usually have dark background as opposed to the bright-field histologically stained images with white background, which can cause a challenge for image transformations due to the lack of pixel-level supervision. To overcome this problem, in addition to inverting the intensities of label-free input images[33,57], other loss terms such as saliency constraint loss[26], and multiscale structural similarity index measure (MS-SSIM) loss[40] were adopted. Although the performance of unsupervised learning is in general inferior to supervised learning[35,57,31,37,48], it still provides a valuable solution in the cases where paired image datasets are not accessible/available for training.

In addition to these mainstream efforts, other novel network architectures and customized loss functions have been recently reported for virtual staining. For example, Liu *et al*. used pathology consistency loss extracted from an additional downstream neural network to guide the training of CycleGANs[46], which achieved staining quality improvements from H&E to Ki-67 stain. As another example, Meng *et al*. demonstrated a parallel feature fusion network (PFFN) that extracts and synthesizes the features from multiscale dimensions to enhance the quality of the virtual H&E images generated from autofluorescence images[28]. In addition, a pyramid pix2pix architecture was exploited by Liu *et al*. to calculate feature losses at multiple spatial scales, which enabled a better transformation from H&E to IHC compared with some of the other popular algorithms[48]. It is anticipated that further improvements in virtual staining quality can be achieved with additional advances in the training of deep learning-based image-to-image transformations.

**Virtual Staining Model Evaluation**

After the training of a virtual staining model, its validity needs to be thoroughly assessed with qualitative and quantitative analysis (Fig. 7). A basic and straightforward assessment method is to directly measure the degree of agreement between the virtually generated histological images and their chemically stained counterparts (ground truth) using standard quantitative metrics (Fig. 7a). When paired input and ground truth images are available, pixel-wise evaluation metrics, such as SSIM[79], peak signal-to-noise ratio (PSNR)[81], MS-SSIM[82], MSE, are MAE are commonly used. When paired input and ground truth images are not available, reference-free metrics can be applied, such as Fréchet inception distance (FID)[83] and inception score (IS)[84], which evaluate the performance of a generative model by measuring the statistical similarity of its output and target images by comparing the high-level features extracted using a trained network[28]. Most of the virtual staining networks developed in the literature used one or more of these standard quantitative metrics to evaluate the image quality of the network inference (see Table 1 and Table 2).

To better assess the model performance within the context of histology, a further step ahead is to extract the critical cellular features from both the virtually stained images and their ground truth images, followed by evaluating the correlations between these extracted features to statistically reveal the level of histology equivalence (Fig. 7b). For example, Bai *et al*. performed color deconvolution and separated different HER2 stain channels[85] to extract the nuclear and membrane features, based on which a good agreement of the statistical signatures between the virtual staining output and ground truth images was found[31]. Likewise, feature-based quantitative analyses performed on segmented Ki-67 positive and negative stained areas[46], segmented cytoplasm and nucleus[51], segmented epithelium and lumen[47], and segmented tumor and stroma regions[53] were reported to validate the virtual staining efficacy of trained network models.

Despite using various image quality metrics and feature analysis tools, algorithmic scores cannot always accurately reflect the diagnostic value of the virtual histology images since the pathologically meaningful features are complex and cannot always be explicitly described through simple numerical rules. Before deployment, it is crucial to validate that the virtually stained images convey the same diagnostically relevant information as the conventional histologically stained slides have. Therefore, including certified pathologists in case studies to assess the important pathological features and make diagnostic decisions from virtually generated images is an important part of the evaluation process (Fig. 7c). For example, in the work of de Haan *et al*.[21], three nephropathologists confirmed that the generation of the additional virtual special stains from existing H&E images improved the diagnosis of kidney diseases; similarly, Bai *et al*.[31] involved three board-certified breast pathologists in validating the diagnosis accuracy and the staining quality of the virtually generated HER2 images. As another example, Lahiani *et al*.[52] included two pathologists in their study to validate the high degree of agreement between the virtually stained images and the corresponding histologically stained ones.

Similar to carrying out a case study with human pathologists evaluating virtually stained images, a digital pathology deep neural network (DNN) model can also be trained to perform multiple types of downstream diagnostic analyses such as cancer stage grading[86] (Fig. 7d). These digital pathology models can create effective evaluation tools in case studies where the clinically relevant features of virtually and histologically stained images are compared against each other. More importantly, these automated image comparison and evaluation tools are fully scalable to be used in large-scale studies, potentially eliminating a bottleneck due to the limited availability of pathologists. For example, Kaza et al.[39] trained a classification model to distinguish the dead and viable cells and their subtypes, which was used to validate the performance of a virtual staining model. Similarly, Li et al.[26] trained a downstream CNN for colonic gland segmentation to demonstrate that the virtually stained images preserve the same rich histopathological information as the histochemically stained ones.

**Discussion and Future Perspectives**

Deep learning-based virtual staining techniques have enabled rapid, cost-effective, and chemical-free histopathology, providing a powerful alternative to the traditional histological staining methods developed over a century. Besides time, cost and labor savings, virtual staining also inherently carries the capability of stain multiplexing. Different types of stains can be simultaneously generated at the same tissue cross-section using a single (or multiple) virtual staining model(s) to provide additional histological information that aids the diagnostic evaluation[21]. This additional histological information was also proven to improve the performance of other downstream machine vision tasks in digital pathology, such as the detection or segmentation of pathological signatures[27,47,50,52,55] and classification of malignancies[32,39,87]. By allowing different stains to be performed on the same tissue section, more tissue will be preserved and be available in diagnostically challenging cases for ancillary tests (e.g., DNA/RNA sequencing) that may be required to reach a diagnosis.

Along with the advancement of this emerging technology, further contributions will be needed to accelerate the development and adoption of virtual staining applications. Such efforts will include the promotion of data consistency, improvement of the staining throughput, incorporation of the latest deep learning advances to improve the generalization of virtual staining networks, and the establishment/validation of better model characterization methods, which are further discussed below.

Like most deep learning-based data-driven techniques, the accessibility of large amounts of high-quality data is the key to successfully training the virtual staining models. However, creating a virtual staining dataset poses unique challenges due to the technical limitations of generating consistent histological ground truth images. The staining results suffer from lab-to-lab and histotechnologist-to-histotechnologist variations, which can be partially attributed to the variations in the protocols and practices in histology labs[88]. During the WSI digitization, the use of different image sensors, objective-lenses and image pre-processing pipelines (e.g., image sharpening, autofocusing, and color correction) among whole slide scanners of different brands makes such variations even worse. Furthermore, the standard histochemical staining procedures also introduce severe mechanical distortions and even damage to the tissue sections, resulting in difficulties in image registration for supervised learning. For example, in Bai *et al.*'s virtual HER2 staining work, approximately 30% of the histochemically stained samples were discarded due to the physical loss of tissue or histochemical stain failures[31]. All factors considered, obtaining high-quality ground truth images can be excessively slow and costly. Therefore, it would be especially beneficial to create large-scale, standardized, and publicly available datasets for the researchers in this field to work with, which will provide a standard testbed for various new methods that are emerging.

Another direction of future research can be on the improvement of the throughput of the virtual staining methods. After several decades of industrial development, the standard histological staining and image digitization process was largely accelerated by automated batch staining and scanning WSI equipment, reaching a high throughput that is necessary to catch up with the clinical needs. On the other hand, some label-free virtual staining methods, though bypassing the chemical staining procedures, employ relatively slow imaging modalities, such as FLIM[34]. Dedicated engineering efforts need to be made for optimized imaging hardware and protocols to enable high-throughput virtual staining methods that can widely replace their standard histological counterparts; for this goal, label-free imaging modalities need to be able to routinely scan/digitize a whole slide image (with a tissue area spanning several $cm^2$) within a few minutes.

Further efforts can boost this virtual staining technology even to surpass the performance of the standard histological staining, and it can potentially be used to virtually stain cellular elements that current methods fail to highlight, such as heavily masked antigens, proteins with low expression levels, and possibly assist in the detection of genomic aberrations (e.g., oncogene amplifications, deletions, and fusions) that require expensive ancillary tests not available in many pathology labs. Moreover, developing fast and stable label-free imaging systems for non-fixed fresh tissue samples and implementing virtual staining on them will be an impactful direction to investigate, which could eliminate the need for biopsies in some anatomical locations and greatly benefit intraoperative consultation during surgical operations.

We anticipate that virtual staining technology will continue to be improved by utilizing state-of-the-art deep learning technologies that are rapidly evolving with more versatile network architectures, new task-specific loss functions, and more efficient training strategies. For example, transformers as an emerging backbone for deep learning tasks have shown superior performance compared to convolution neural networks on various computer-vision tasks[89–91], which might also provide promising improvements for virtual staining networks, potentially offering higher image resolution and staining accuracy. Additionally, loss functions specifically designed for pathological images using handcrafted expert features or pre-trained feature extractors could be introduced during the training phase to act as domain-specific penalty terms that might improve the generalization of virtual staining networks. Other emerging training strategies, such as learning rate scheduling, large-scale parallel training, and different normalization schemes, will also be important to further advance the capabilities of virtual staining networks.

Despite the promising technical feasibility and proof-of-concept demonstrations summarized in this Review, the implementation of virtual staining technology for primary diagnostic use in clinical settings is yet to arrive (which will need to go through a Class III approval process through the FDA). For this, the accuracy and reliability of the virtual staining technology need to be fully characterized/validated by different medical institutions using a broad distribution of tissue samples from a large number of patients with diverse pathologies. To relieve potential concerns regarding virtual staining network hallucinations, various quantitative metrics were developed (see the Model Evaluation section discussed earlier), which can be used to assess the model efficacy and the image quality of the virtual staining outcomes. Based on these existing metrics, a quantitative benchmark for the clinical success of virtual staining technology needs to be established to reflect the level of diagnostic errors or uncertainties due to the chemical staining and interobserver variabilities, which will provide a reference for all the virtual staining studies to compare with and guide the proper design of case studies. In addition, the virtual staining technology development and advancement phase will need fast and quantitative feedback during the iterative development procedures to converge on competitive models that can be tested in clinical settings. Developing automated and reliable evaluation tools (such as a set of task-specific neural networks) that can partially replace human diagnosticians or pathologists during this research and development phase will greatly accelerate the virtual staining research progress since the availability of well-trained pathologists for large-scale, multi-institution validation efforts might introduce challenges. Such automated and repeatable image quality evaluation tools will also be valuable in fostering the design of large-scale case validation studies at a global scale and help accelerate the clinical acceptance of virtual staining techniques.

## Table 1. Label-free virtual staining studies using deep learning

| Authors | Input (Label-free imaging modalities) | Output (Stain types) | Organs | Paired/ Unpaired | Model evaluations | | | |
|---|---|---|---|---|---|---|---|---|
| | | | | | Standard quantitative metrics | Algorithm-based feature analysis | Learning-based downstream analysis | Pathologists' assessments |
| Rivenson et al.[18,25] (2018) | Autofluorescence | H&E | Salivary gland, thyroid | P | | | | ✓ |
| | | Jones | Liver, lung, kidney | P | | | | |
| | | MT | Liver, lung, kidney | P | | | | |
| Rivenson et al.[20,92] (2018) | Quantitative phase | H&E | Skin | P | ✓ | | | |
| | | Jones | Kidney | P | | | | |
| | | MT | Liver | P | | | | |
| Borhani et al.[34] (2019) | TPEF + FLIM | H&E | Rat liver | P | ✓ | | | |
| Nygate et al.[32] (2020) | Quantitative phase | QuickStain | Sperm cells | P | ✓ | | | ✓ |
| Zhang et al.[22] (2020) | Autofluorescence + Stain matrix | H&E / MT / Jones | Kidney | P | ✓ | | | |
| Li et al.[36] (2020) | Bright-field | H&E / PSR / Orcein | Rat carotid artery | P | | | | ✓ |
| Pradhan et al.[35] (2021) | Nonlinear multi-modal (CARS+TPEF+SHG) | H&E | Colon | P, U | ✓ | | | |
| Li et al.[26] (2021) | Autofluorescence | H&E | Colon | U | ✓ | | ✓ | ✓ |
| Picon et al.[27] (2021) | Autofluorescence (Lissajous-subsampled) | H&E | Colon, breast, lung | P | ✓ | | ✓ | |
| Meng et al.[28] (2021) | Autofluorescence | H&E | Ovarian | U | ✓ | | | ✓ |
| Li et al.[23] (2021) | Reflectance confocal microscopy | Acetic acid stain / H&E | Skin | P | ✓ | ✓ | | |
| Bai et al.[31,93] (2021) | Autofluorescence | HER2 IHC | Breast | P | ✓ | ✓ | | ✓ |
| Kang et al.[40] (2022) | Ultraviolet photoacoustic microscopy | H&E | Mouse brain | U | ✓ | ✓ | | |
| Zhang et al.[37] (2022) | Bright-field | H&E / PSR / EVG | Carotid artery | P | ✓ | | | ✓ |
| Kaza et al.[39] (2022) | UV imaging | Giemsa staining | Blood Smears | P | ✓ | | ✓ | |
| Boktor et al.[38] (2022) | Total-absorption photoacoustic remote sensing | H&E | Skin | P | ✓ | | | ✓ |
| Abraham et al.[33] (2022) | Quantitative oblique back-illumination microscopy | H&E | Rat brain | U | | | | |
| Cao et al.[41] (2022) | Ultraviolet photoacoustic microscopy | H&E | Bone | U | | ✓ | | ✓ |

**Staining abbreviations: EVG**: Verhoeff's Van Gieson

| Table 2. Stain-to-stain transformation studies using deep learning | | | | | | | | |
|---|---|---|---|---|---|---|---|---|
| **Authors** | **Input (Stain types)** | **Output (Stain types)** | **Organs** | **Paired/ Unpaired** | **Model evaluations** | | | |
| | | | | | Standard quantitative metrics | Algorithm-based feature analysis | Learning-based downstream analysis | Pathologists' assessments |
| Gadermayr et al.[43] (2018) | PAS | AFOG / Col3 / CD31 | Kidney | U | | | | |
| | AFOG / Col3 / CD31 | PAS | | | | | | |
| Levy et al.[51] (2020) | H&E | Trichrome | Liver | U | | | | ✓ |
| | | SOX10 IHC | Skin, lymph node | P, U | | ✓ | | |
| Mercan et al.[50] (2020) | H&E | PHH3 IHC | Breast | U | | | ✓ | |
| | PHH3 IHC | H&E | | | | | | |
| de Haan et al.[21,94] (2020) | H&E | Jones / MT / PAS | Kidney | P | | | | ✓ |
| Burlingame et al.[56] (2020) | H&E | PanCK IF | Pancreas | P | ✓ | | ✓ | |
| Lahiani et al.[52] (2021) | H&E | FAP-CK IHC | Liver | U | ✓ | | | ✓ |
| Liu et al.[46] (2021) | H&E | Ki-67 IHC | Neuroendocrine tumor, breast | U | ✓ | ✓ | | |
| Hong et al.[53] (2021) | H&E | Cytokeratin IHC | Stomach | P | | ✓ | | ✓ |
| Chen et al.[57] (2021) | Hoechst stained MUSE images | H&E | Mouse brain, mouse liver | U | ✓ | | ✓ | |
| Ghahremani et al.[55] (2022) | IHC | Multiplex IF | Lung, bladder, breast, colon, prostate | P | ✓ | ✓ | ✓ | ✓ |
| Xie et al.[47] (2022) | H&E | CK8 IHC | Prostate | P | ✓ | ✓ | | ✓ |
| Zhang et al.[49] (2022) | H&E | Ki67 / CC10 / proSPC IHC | Mouse lung | U | ✓ | ✓ | | ✓ |
| | H&E | ER / PR / HER2 IHC | Breast | | | | | |
| | H&E | Oil red O / a-SMA IHC / macrophages | Rabbit cardiovascular | | | | | |
| Bouteldja et al.[95] (2022) | IHC: CD31 / Col3 / NGAL / a-SMA | PAS | Kidney | U | | | ✓ | |
| Lin et al.[45] (2022) | H&E | MT / PAS / PASM | Kidney | U | ✓ | | | |
| Yang et al.[76] (2022) | H&E | PAS | Kidney | P | ✓ | ✓ | | |
| Liu et al.[48] (2022) | H&E | HER2 IHC | Breast | P | ✓ | | | ✓ |

**Staining abbreviations: FAP-CK**: Fibroblast Activation Protein and Cytokeratin, **CC10**: Clara cell 10, **proSPC**: Anti-Prosurfactant Protein C, **ER**: Estrogen receptor, **PR**: Progesterone receptor, **PASM**: Periodic Schiff-Methenamine

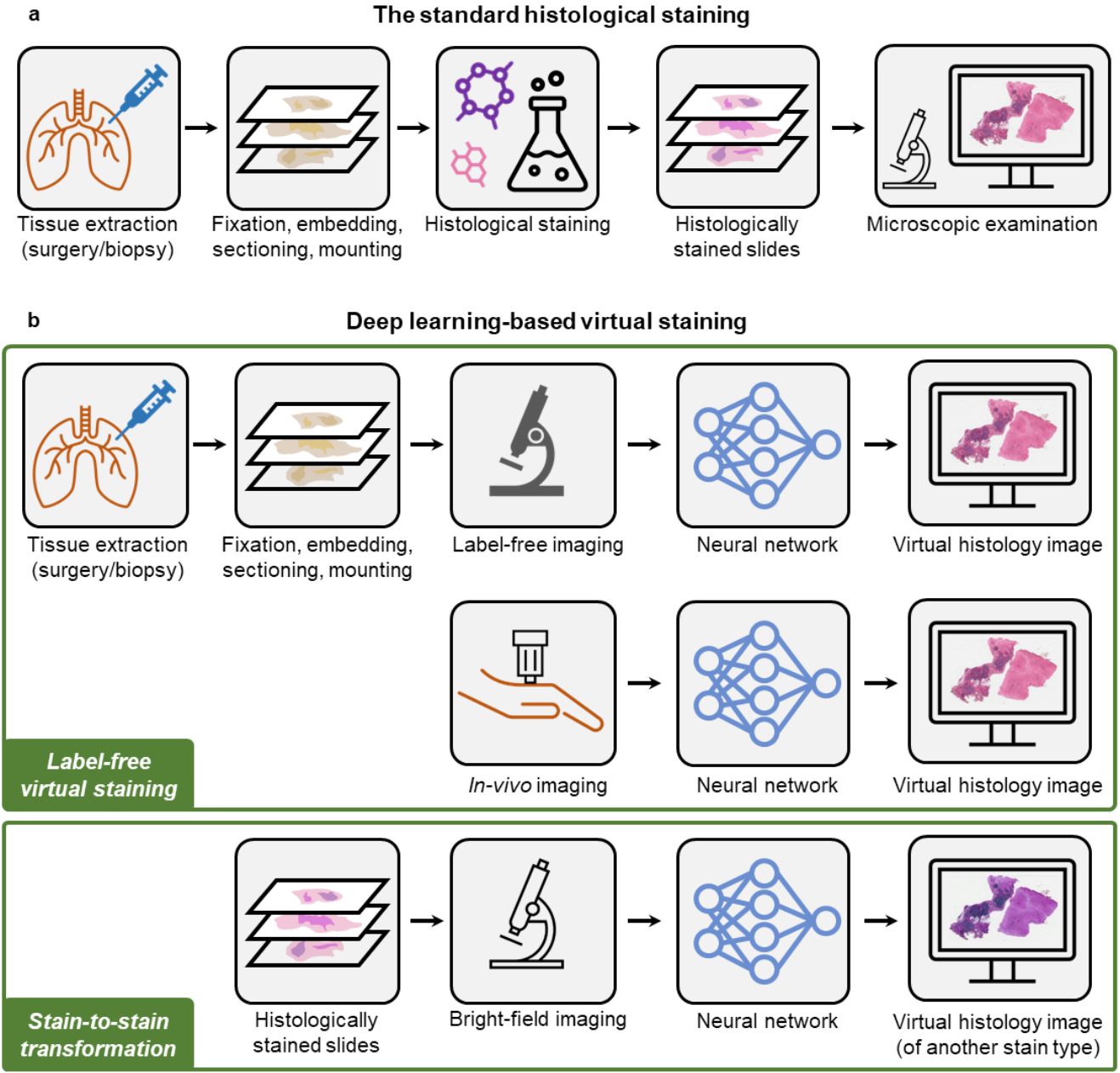

**Figure 1. Schematic of the standard histological staining and deep learning-based virtual staining.**
(**a**) Standard histological staining relies on laborious chemical-based tissue processing and labeling steps.
(**b**) Pre-trained deep neural networks enable the virtual histological staining of label-free samples as well as the transformation from one stain type to another, without requiring any additional chemical staining procedures.

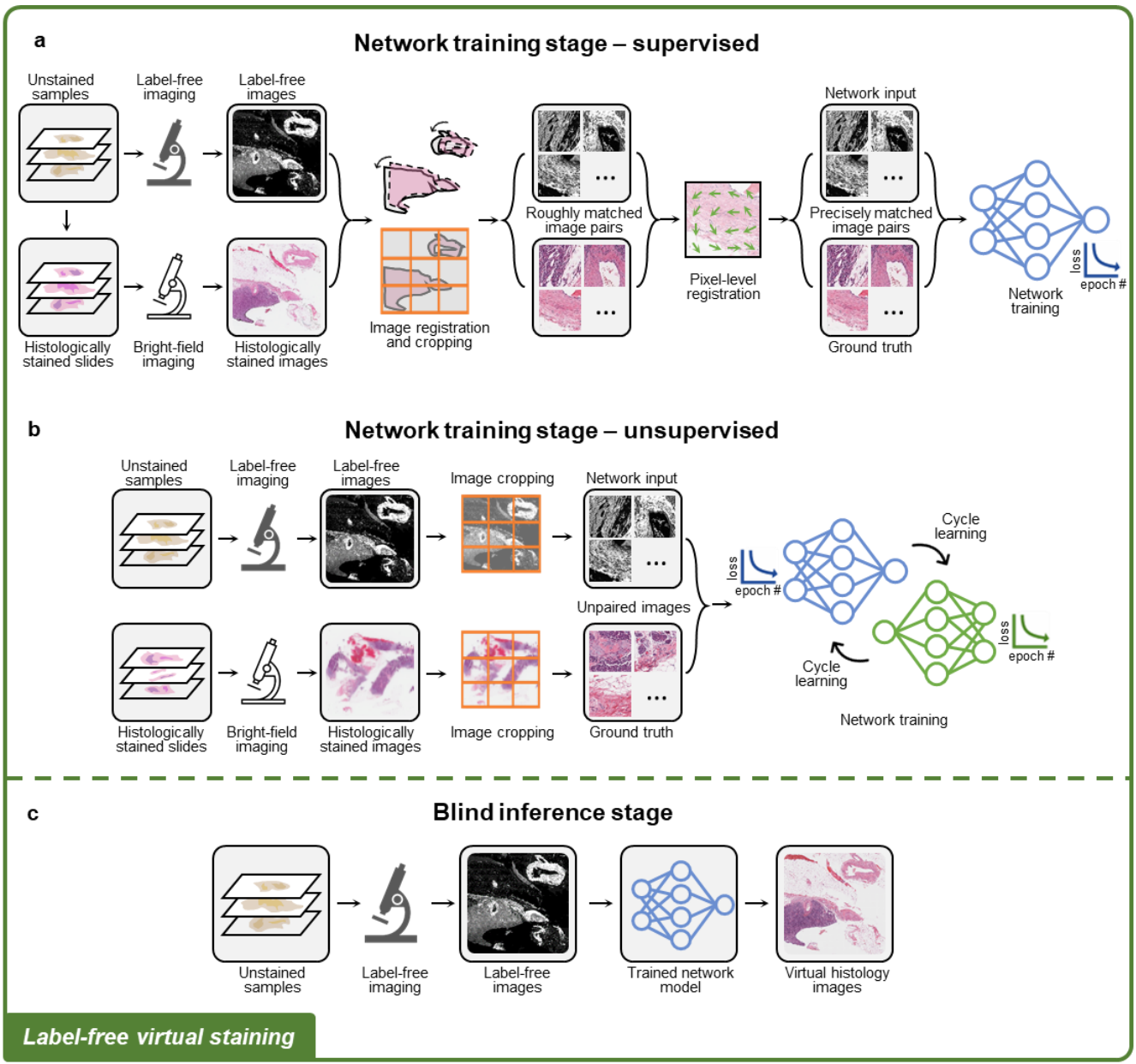

**Figure 2. Training and inference of label-free virtual staining networks. (a)** Training of a label-free virtual staining network using the supervised scheme. Precisely matched input and ground truth image pairs are required, which can be obtained through a multi-stage image registration process. **(b)** Training of a label-free virtual staining network using the unsupervised scheme, in which input and ground truth images are not necessarily paired. Cycle-consistency-based learning frameworks are typically used. **(c)** Blind inference of a trained virtual staining model. The virtual histology images are rapidly generated from label-free images using a digital computer.

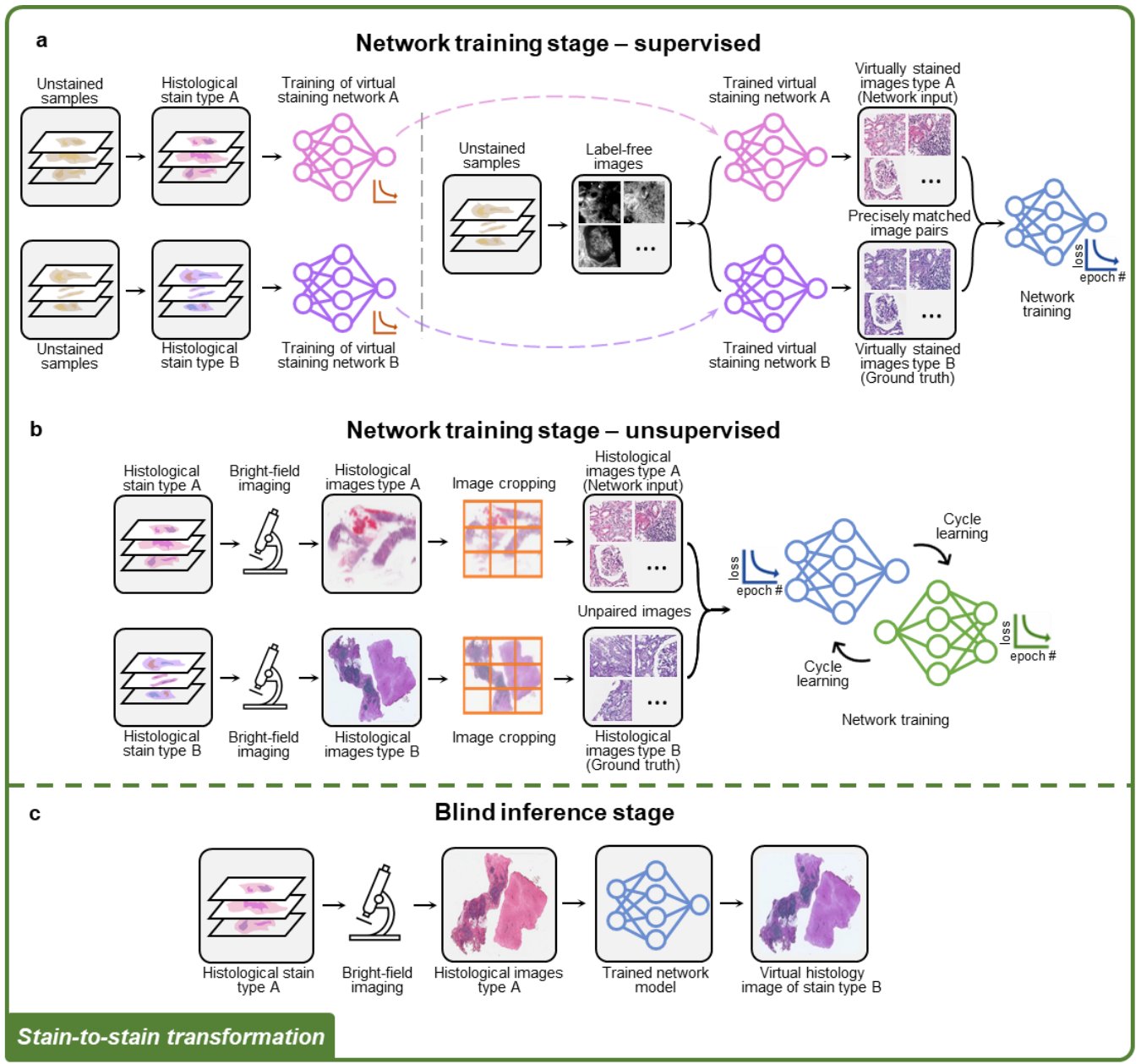

**Figure 3. Training and inference of stain-to-stain transformation networks. (a)** Training of a stain-to-stain transformation network using the supervised scheme. **(b)** Training of a stain-to-stain transformation network using the unsupervised scheme, in which input and ground truth images are not necessarily paired. **(c)** Blind inference of a trained stain-to-stain transformation model. Additional histological stain types can be generated from the existing stain, providing additional diagnostic information without altering the current histopathology workflow.

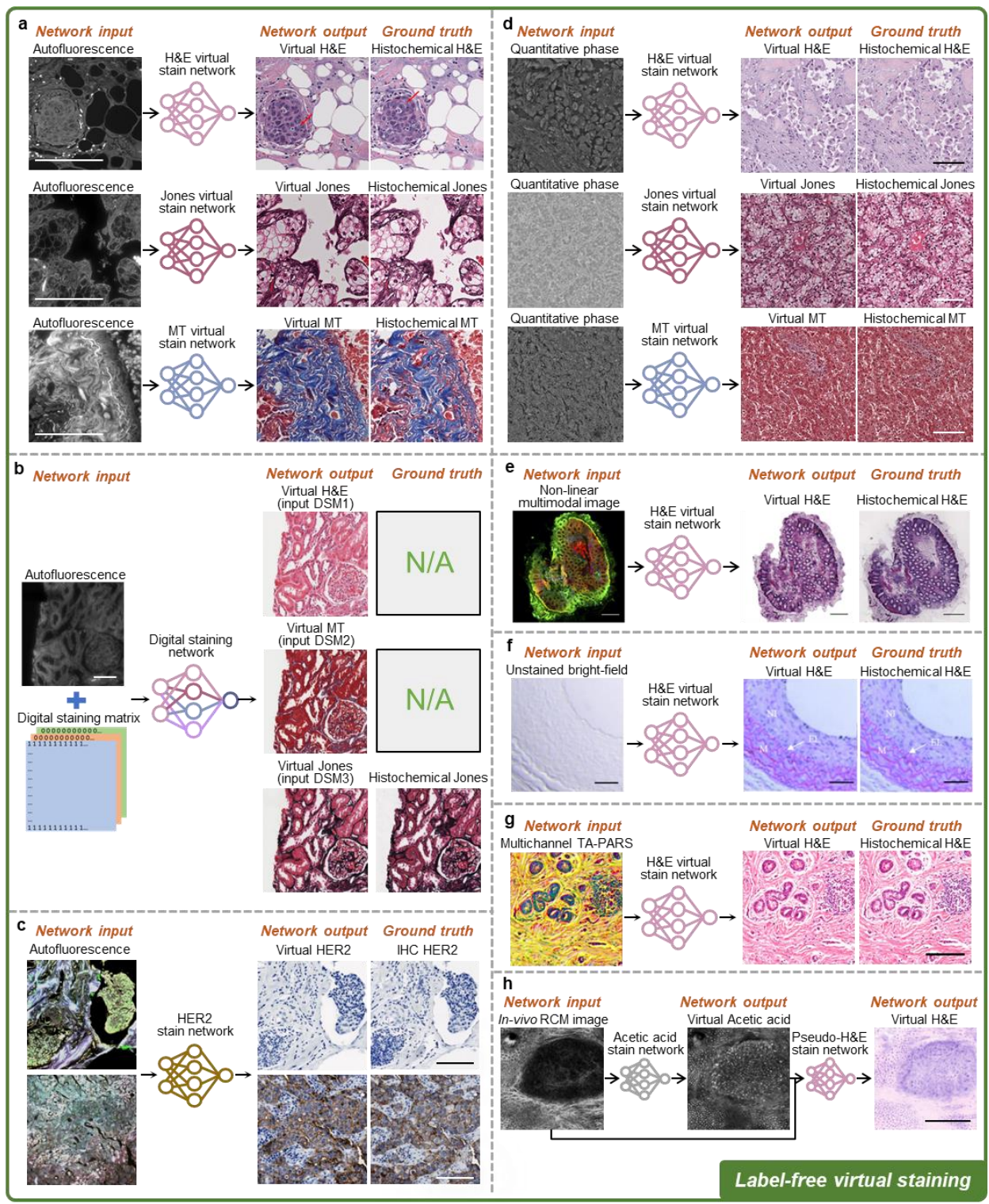

**Figure 4. Examples of label-free virtual staining using different input imaging modalities. (a)** Virtual H&E, Jones silver, and MT staining using autofluorescence images[18]. **(b)** Multiplexed H&E, Jones silver, and MT staining using a single network with autofluorescence images and digital staining matrix as input[22]. **(c)** Virtual IHC HER2 staining using autofluorescence images[31]. **(d)** Virtual H&E, Jones silver,

and MT staining using quantitative phase images (QPI)[20]. **(e)** Virtual H&E staining using nonlinear multimodal images[35]. **(f)** Virtual H&E staining using bright-field images[36]. **(g)** Virtual H&E staining using TA-PARS images[38]. **(g)** Virtual acetic acid and H&E staining using *in vivo* RCM images[38]. All the scale bars represent 100 μm.

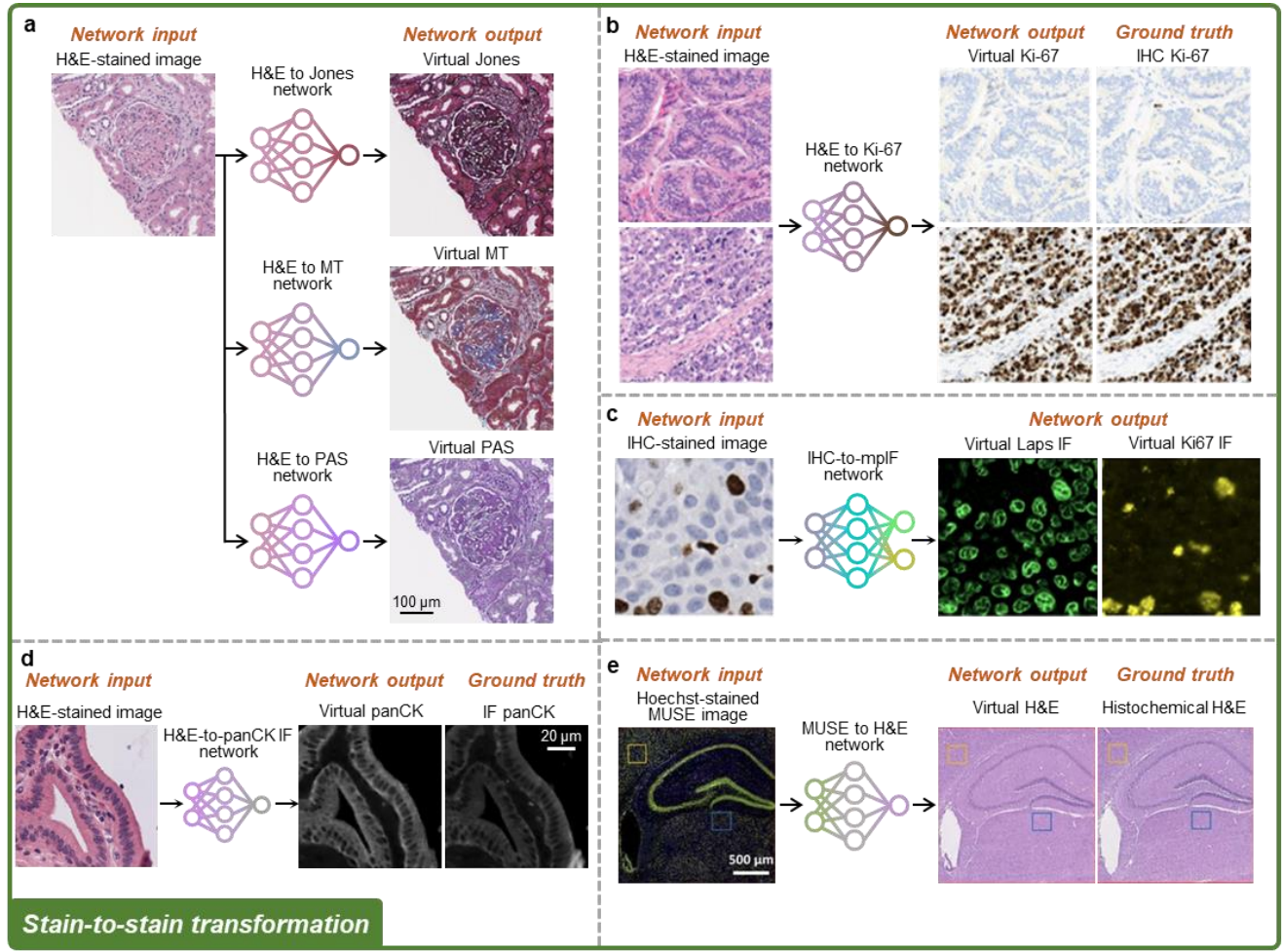

**Figure 5. Examples of virtual stain-to-stain transformations. (a)** Transformation from H&E staining into virtual Jones silver, MT, and PAS staining[21]. **(b)** Transformation from H&E staining into virtual IHC Ki-67 staining[46]. **(c)** Transformation from Ki-67 IHC staining into multiplexed virtual IF staining[55]. **(d)** Transformation from H&E staining into virtual panCK IF staining[56]. **(e)** Virtual H&E staining using Hoechst-stained MUSE images[57].

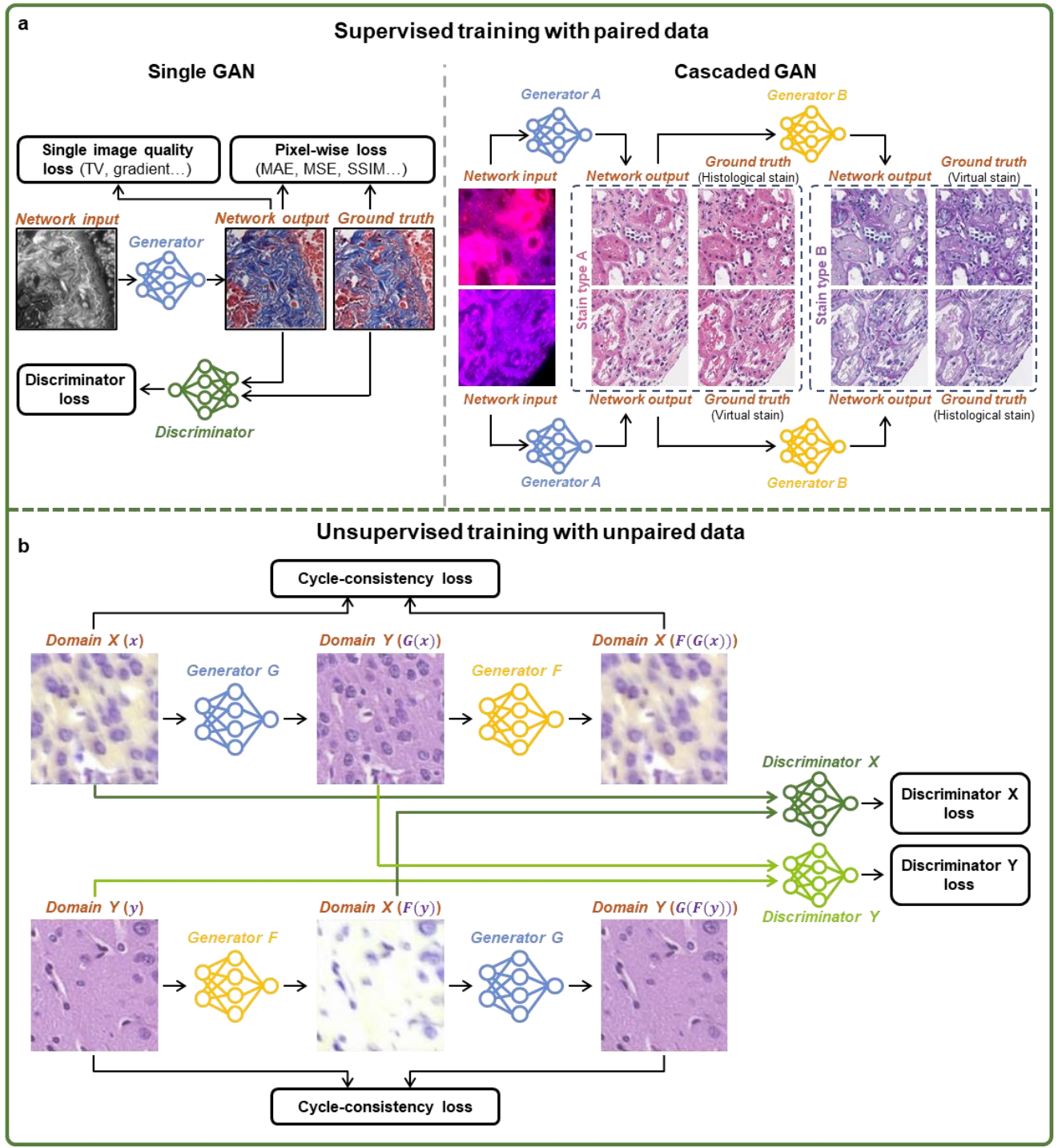

**Figure 6. Virtual staining network architectures and training scheme. (a)** For supervised learning, when paired image data are available, GAN and its variants are typically used. When partially paired data are available, a cascaded GAN that optimizes sequential image transformation models can be used. **(b)** For unsupervised training with unpaired image data, CycleGAN and its variants are typically used.

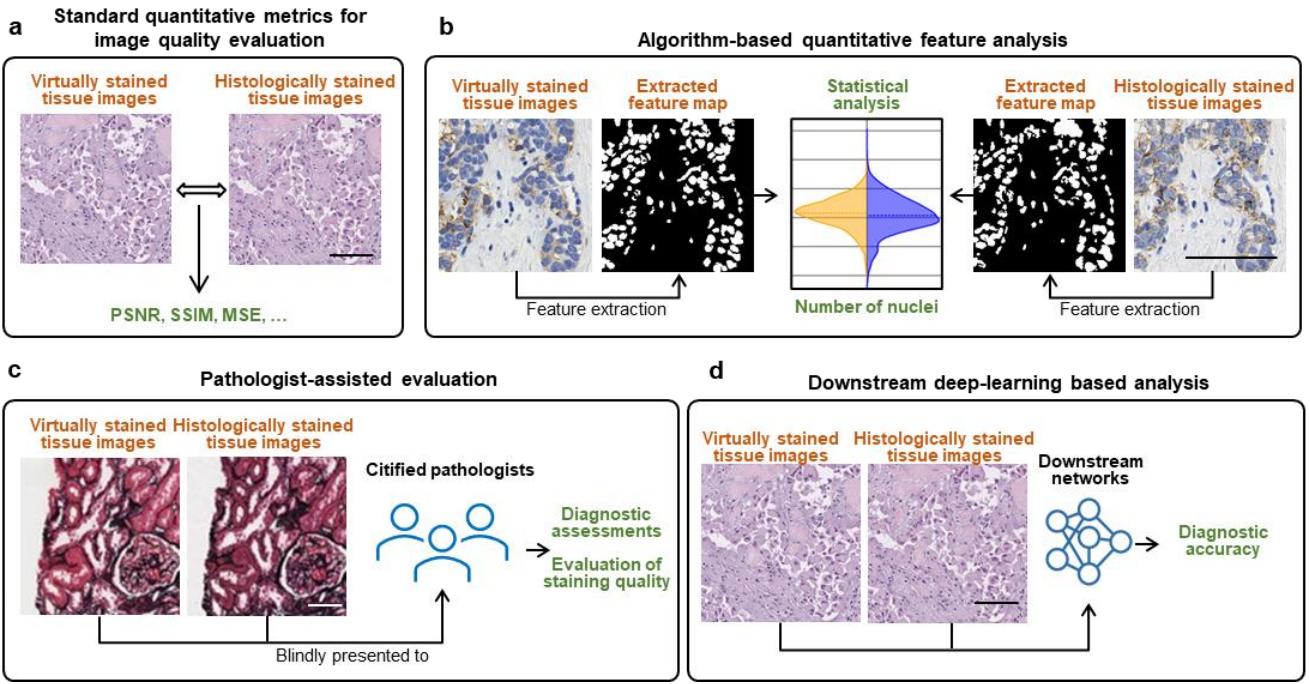

**Figure 7. Evaluation methods for virtual staining neural network models. (a)** Standard quantitative metrics such as PSNR and SSIM are calculated based on output images and their corresponding ground truth images. **(b)** Pathological features are extracted, and the statistical correlations between the features from the virtual and histological staining methods are compared. **(c)** The diagnostic values and the staining quality of the virtually generated images are evaluated by expert pathologists and compared against the histologically stained ones. **(d)** Validated digital pathology DNN models that can perform downstream diagnostic analysis are used to evaluate the clinically relevant characteristics of virtually and histologically stained images.